\documentclass[12pt]{article}
\usepackage{graphicx}
\usepackage{amsmath,amssymb}

\makeatletter \@addtoreset{equation}{section}

\newcommand{\be}{\begin{equation}}
\newcommand{\ee}{\end{equation}}
\newcommand{\bear}{\begin{eqnarray}}
\newcommand{\eear}{\end{eqnarray}}
\newcommand{\ba}{\begin{array}}
\newcommand{\ea}{\end{array}}

\newcommand{\nn}{\nonumber \\}

\newcommand{\tr}{{\rm tr}}

\newcommand{\e}{{\rm e}}

\newcommand{\LS}{\ \ \ \ \ \ \ \ \ \ }

\newcommand{\bsubeq}{\begin{subequations}}
\newcommand{\esubeq}{\end{subequations}}

\newcommand{\w}{\wedge}
\renewcommand{\d}{{\rm d}}

\DeclareFontFamily{U}{rsf}{}
\DeclareFontShape{U}{rsf}{m}{n}{
  <5> <6> rsfs5 <7> <8> <9> rsfs7 <10-> rsfs10}{}
\DeclareMathAlphabet\Scr{U}{rsf}{m}{n}

\begin{document}

\allowdisplaybreaks{

\begin{titlepage}
\vfill
\begin{flushright}
{\tt\normalsize KIAS-P06029}\\
{\tt\normalsize hep-th/0607091}\\
\end{flushright}

\vfill
\begin{center}
{\Large\bf A Heterotic Flux Background \\ and Calibrated  Five-Branes  }

\vskip 0.5in

{ Seok Kim\footnote{\tt seok@kias.re.kr} and Piljin Yi\footnote{{\tt
piljin@kias.re.kr}} }

\vskip 0.15in

{\it School of Physics, Korea Institute for Advanced Study,} \\
{\it 207-43, Cheongryangri-Dong, Dongdaemun-Gu, Seoul 130-722,
Korea}
\\[0.3in]


\end{center}

\vfill

\begin{abstract}
\normalsize\noindent We consider, in flux compactification of
heterotic string theory, spacetime-filling five-branes.
Stabilizing the fivebrane involves minimizing the combined energy
density of the tension and a Coulomb potential associated with an
internal 2-dimensional wrapping. After reviewing the generalized
calibration under such circumstances, we consider a particular
internal manifold based on a $T^2$ bundle over a conformally
rescaled $K3$. Here, we find two distinct types of wrapping. In
one class, the fivebrane wraps the fibre $T^2$ which belongs to
a cyclic homotopy group. The winding number is not extensive, yet
it maps to D3-brane number under a U-duality map to type IIB side.
We justify this by comparing properties of the two sides in detail.
Fivebranes may also wrap a topological 2-cycle of K3, by
saturating a standard calibration requirement with respect to a
closed K\"ahler 2-form $J_{K3}$ of $K3$. We close with detailed
discussion on F-theory dual of these objects and related issues.
\end{abstract}

\vfill

\end{titlepage}
\setcounter{footnote}{0}
\baselineskip 18pt

\section{Introduction}

Over the last few years, flux compactification of string theory has
proven to be a rich playing ground for connecting to real world.
From building a realistic particle physics model to understanding
the inflation era, it has given many new insights. Existence of
landscape
\cite{Susskind:2003kw,Douglas:2003um,Ashok:2003gk,Denef:2004ze},
namely a large number of discrete vacua, stable and semi-stable
\cite{Kachru:2003aw}, offers us a completely different view on our
universe.

Much of these developments came from a special subclass of type IIB
flux compactification, where the internal manifold is a warped
Calabi-Yau \cite{Giddings:2001yu}. General supersymmetry
requirements in type IIB and IIA are known to demand much less,
known as $SU(3)$-structure
\cite{LopesCardoso:2002hd,Gauntlett:2003cy,Gualtieri:2003dx,
Grana:2004bg, Grana:2004sv,Grana:2005sn,Grana:2005jc}, and we do not
yet have a clear picture of vacua in this more general setting.
Notable exceptions are the so-called F-theories, which are IIB
compactifications with nonuniform dilaton-axion. In particular, the
classic example of F-theory on $K3\times K3$ has been explored from
early on with fluxes turned on
\cite{Becker:1996gj,Dasgupta:1999ss,Becker:2002sx,Becker:2003yv,Becker:2003sh,
Aspinwall:2005ad,Becker:2005nb}. This example is also interesting
because of a known U-dual map to the heterotic side.

Flux compactification of the heterotic string theory
\cite{Gross:1984dd} was first considered in 1986 by Strominger, who
gave a complete characterization of supersymmetry requirements
\cite{Strominger:1986uh}. While the geometry is not as simple as
warped Calabi-Yau, it still has an $SU(3)$ holonomy with respect to
a torsionful connection, which is still simpler than the general
$SU(3)$ structure manifold. The heterotic flux compactification is
known to evade the usual no-go theorem against smooth flux
compactification, at the cost of introducing higher order curvature
term in the equation of motion and a Bianchi identity
\cite{Green:1984sg,Bergshoeff:1989de}. It is known that
the smooth compactification is available only if one
stays far away from minimal embedding configuration of the gauge
bundle on the internal manifold \cite{Kimura:2006af,Ivanov:2000fg}.

No explicit solution to the heterotic system is available.
However, Fu and Yau proved an existence theorem for a smooth
solution whose internal manifold is a $T^2$ fibre bundle over the
base of conformally rescaled $K3$ \cite{Fu:2006vj}. It is expected
that this class of solution would map to the above mentioned
$K3\times K3$ F-theory (or its IIB orientifold limit), under a
chain of U-duality map \cite{Becker:2006et}. For the first time,
we have a reasonably explicit dual pair of flux compactification
model which deviates significantly from the conformally Calabi-Yau
examples of IIB, on which much of recent applications are based.
The purpose of this paper is to explore this pair, largely from
the heterotic side, with emphasis on putting extra structures due
to fivebranes. Under the U-dual map from F/IIB side, D-branes must
be emulated by fivebranes, strings, and the gauge bundle. In this
respect, understanding of fivebrane in the heterotic side remains
an important issue in further exploration of this class of
solution.

This note is organized as follows. After a quick summary of
supersymmetry condition with flux in section 2, we import the
generalized calibration of M5 branes and adapt it to the heterotic
string theory in section 3. Here flux acts to contribute magnetic
Coulomb energy to fivebranes which modifies the problem of finding
supersymmetric brane configurations significantly. Section 4
outlines Fu-Yau solution and delineates how the generalized
calibration of fivebrane specializes to this background. Two types
of spacetime-filling fivebranes are found. One that winds around a
homotopically trivial $T^2$ fibre, and those which wrap certain
homology 2-cycles and orthogonal to the flux. Section 5 and 6 are
devoted to complete characterization of these fivebranes,
including detailed issue of tadpole conditions. In particular,
fivebranes winding on $T^2$ are identified as U-dual of D3 branes
in IIB or F-theory, despite the fact that the winding number is
not an extensive quantity. We show that the cyclic nature of the
$T^2$ winding number is in fact precisely mirrored in the number of
D3 branes on F/IIB side. Section 7 tidies up
some loose ends associated with this U-dual map to F/IIB side, where
among other things, counting of worldvolume degrees of freedom are
considered with and without fluxes.

\section{Flux Compactification of Heterotic String Theory}

We start with the bosonic part of the
supergravity/super-Yang-Mills action in ten
dimensions,\footnote{We follow \cite{Polchinski} for conventions,
except for using anti-hermitian basis for the gauge field.}
\begin{align}\label{action}
\begin{split}
\Scr{L} \ &= \
\frac{1}{2\kappa_{10}^2} \sqrt{- G} \, \e^{-2 \Phi}
\Bigg[
R (\omega) - \frac{1}{12} H_{MNP} H^{MNP}
+ 4 (\nabla_M \Phi)^2
\\
\ & \LS \LS \ \
+ \frac{\alpha'}{8} \Big\{ \tr (F_{MN} F^{MN}) - \tr (R_{MN} (\omega_-)
R^{MN} (\omega_-)) \Big\}
\Bigg]
\; ,
\end{split}
\end{align}
with anti-Hermitian basis for the gauge field and anti-symmetric
generators for the curvature, both of the unit normalization.
According to Bergshoeff et.al. \cite{Bergshoeff:1989de}, the
curvature that appears in the last term of the action is the one
with torsion $-\frac{1}{2}H$ with the convention that the
connection with torsion $+\frac{1}{2}H$ appears in the
supersymmetry variation of gravitino.

With supersymmetry, the metric in string frame has
no warp factor,
\begin{align}
G_{MN} \, \d x^M \d x^N
\ = \
\eta_{\mu \nu} \, \d x^{\mu} \d x^{\nu}
+ g_{mn} \, \d y^m \d y^n \; ,
\end{align}
with a metric $g_{mn}$ on the compact manifold ${\cal M}_6$.
Supersymmetry also implies  a complex structure $J$ which is
integrable,
\begin{align}
0 \ = \ N_{mn}{}^p \ &= \
J_m{}^q \nabla_{[q} J_{n]}{}^p - J_n{}^q \nabla_{[q} J_{m]}{}^p
\; ,
\end{align}
with respect to which the metric $g_{mn}$ is hermitian.
$J$ is also parallel under a torsionful connection
\begin{align}
\nabla_m^{(+)} J_{np} \ &= \ 0 \; ,
\; .
\end{align}
with the connection having a torsion $H/2$.\footnote{That is, the
spin connection is shifted as $\omega^a_{\;\;b\mu}\rightarrow
\omega^a_{(+)b\mu}=\omega^a_{\;\;b\mu}+\frac{1}{2}H^a_{\;\;b\mu}$
in our normalization.}

Supersymmetry relates the gradient of the complex structure
$J$ and that of the dilaton $\Phi$, and the antisymmetric tensor $H$.
First, $H$ can be
identified with the so-called Bismut torsion \cite{Bismut89,LopesCardoso:2003af}
\begin{equation}
H_{mnp} \ = \ -
3 J_m{}^q J_n{}^r J_p{}^s \nabla_{[q} J_{rs]} \; ,
\end{equation}
and the dilaton is related to $J$ as
\begin{equation}
\nabla_m \Phi \ = \
\frac{3}{4}J^{np} \nabla_{[m}J_{np]}
\; .
\end{equation}
The relation between dilaton and $H$ can be also read off from the
above,
\begin{align}
\nabla_m \Phi \ = \  \frac{1}{4} J_{mn}J_{pq}H^{npq} \; ,
\end{align}
and tells us that the non-primitive part of $H$ is
fully encoded in $\d\Phi$.

Note that, from these, we also have
\begin{equation}
0=\d \left(e^{-2\Phi} J \w J \right) \; ,
\end{equation}
and
\begin{equation}
H=- * e^{2\Phi}\d \left(e^{-2\Phi} J\right) \; ,
\end{equation}
once we make use of the hermiticity. Throughout this note, $*$ means
the Hodge-star operation with respect to the six dimensional metric
$g$. We will always add subscripts to signify Hodge-star operations
with respect to other metrics. In particular, the latter can be
rewritten as
\begin{equation}
*\left(e^{-2\Phi}H\right)=\d \left(e^{-2\Phi} J\right) \; ,
\end{equation}
which automatically solves the equation of motion for $H$.

In compactifying the heterotic theory, an important topological
constraint is found in the Bianchi identity of $H$, which reads,
 \begin{align}
\d H = \frac{\alpha'}{4} \Big[ \tr ( R'  \w R' )
- \tr (F \w F) \Big]
\; ,
\end{align}
The curvature 2-form $R'$ is ambiguous as far as anomaly
cancelation is concerned \cite{Hull:1985dx}. While the natural
choice with the given form of the action would be that of the
connection with torsion $-H/2$
\cite{Hull:1986kz,Bergshoeff:1989de}, any shift of torsion piece
is allowed \cite{Sen:1986mg}. Also, as elaborated in
Ref.\cite{Kimura:2006af}, a smooth and large compactification
necessarily implies that $H^2$ is of order $\alpha'$ so the
torsion part of $R'$ contributes only a higher order term on the
right hand side. A convenient choice for $R'$ is to take the
so-called Hermitian connection, whereby the right hand side is of
Hodge type $(2,2)$. Since $\d H$ is of Hodge type $(2,2)$ also,
this choice represents a consistent truncation of this equation.

When fivebranes are present, the Bianchi identity will acquire
source terms as
 \begin{align}
\d H =\frac{ \alpha'}{4} \Big[ \tr ( R'  \w R'  )
- \tr (F \w F)  - 16\pi^2\delta_{fivebrane}\Big]
\; ,
\end{align}
which could modify the topological constraint on the gauge bundle.
Another way to view this is to consider the fivebranes as singular
limits of the gauge bundle, where the Hermitian Yang-Mills
degenerates such that $\tr F\w F$ becomes a delta function source.
The convention here is such that supersymmetric $F$ is
anti-self-dual. Since we are using the anti-hermitian basis with
unit normalization also, the density $ tr F\wedge F/16\pi^2$
integrates to a nonnegative integer against a topological 4-cycle.

\section{Generalized Calibration for Fivebranes}

Let us consider fivebranes which are spacetime-filling in a
heterotic flux compactification with a $3+1$ Minkowski spacetime
intact. Such fivebranes may be alternatively considered as small
instanton limits of the gauge bundle. They would wrap two-cycles in
internal manifold and thus are of co-dimension 4 objects.  The two
worldvolume directions along ${\cal M}_6$ should span out a
two-dimensional surface, which we will denote by $\Sigma$. In this
note, we will consider such fivebranes in the probe limit but will
be careful to maintain the tadpole cancelation conditions for
consistency.

For simplicity, we will consider configurations with no worldvolume
field strength turned on. These are completely characterized by the
two-dimensional embedding $\Sigma$ into ${\cal M}_6$, and we are
lead to the following energy functional per unit 3-volume,
\begin{equation}
vol_{3+1}\;\lrcorner\;\left[\int_\Sigma e^{-2\Phi}  vol_{5+1} -
\int_{\Sigma}  B_6\right] \;,
\end{equation}
to be minimized, where $B_6$ is a 6-form potential dual to $B_2$
and $ \lrcorner$ denotes exterior contraction between differential
forms. The first term is the contribution from the tension of the
fivebrane, which explains the presence of $e^{-2\Phi}$, while the
second is the minimal coupling which incorporates the energy due
to Coulombic potential $B_{6}$. With the above general form of
$H$, the dual potential is computed as
\begin{equation}
dB_6= e^{-2\Phi}*_{9+1}H= e^{-2\Phi}(* H)\w vol_{3+1}=d(e^{-2\Phi}J)\wedge vol_{3+1} \;,
\end{equation}
which gives
\begin{equation}
B_6=e^{-2\Phi}J\w vol_{3+1} \;,
\end{equation}
up to an additive ambiguity of a closed 6-form. Since
\begin{equation}
vol_{5+1}=vol_\Sigma\w vol_{3+1} \;,
\end{equation}
the energy functional of the wrapped fivebrane, per unit 3-volume, is
\begin{equation}
{\cal E}(\Sigma)=\int_\Sigma e^{-2\Phi}vol_\Sigma-\int\Sigma^*(e^{-2\Phi}J)\; .
\end{equation}
A stable configuration of a wrapped fivebrane is obtained only if ${\cal E}$
is minimized against deformation of the embedding $\Sigma$ into ${\cal M}_6$.
\begin{equation}
\delta{\cal E}(\Sigma)= 0 \; .
\end{equation}
With the current gauge choice of $B_6$, in particular, this energy
functional is clearly nonnegative. One class of minimized solution
would be obtained for $\Sigma$'s for which we have
\begin{equation}
{\cal E}(\Sigma)=0 \; .
\end{equation}
Here we should emphasize that the value of ${\cal E}$ is not
necessarily the physical energy. Only its variation is important
for our purpose. In a later example, we will shift the definition
of the energy by a closed form, giving it a more familiar shape.

Minimization of ${\cal E}$ is clearly a necessary condition for
supersymmetry. Here we would like to show that ${\cal E}=0$ is a
consequence of generalized calibration; in other words,
configurations with ${\cal E}=0$  are actually supersymmetric. For
this, we will regard our fivebranes as M5 branes in the heterotic
M-theory \cite{Duff:1995wd,Horava:1995qa,Horava:1996ma}, which are
orthogonal to the 11-th direction. The metric of the spacetime is
\begin{eqnarray}
G^{10+1}_{AB}\d X^A\d X^B&=&e^{4\Phi/3}\d x_{11}^2+e^{-2\Phi/3}G_{MN}\d x^M\d x^N \nn
&=& e^{4\Phi/3}\d x_{11}^2+e^{-2\Phi/3}\left(\eta_{\mu\nu}
\d x^\mu \d x^\nu+g_{mn}\d y^m \d y^n\right) \; ,
\end{eqnarray}
for an interval along $x_{11}$,
while there is nontrivial 3-form potential $C_3$
\begin{equation}
\d C_3= dx_{11}\wedge \frac16 H_{mnp}\,\d y^m \w\d y^n\w \d y^p \; ,
\end{equation}
whose equation of motion give us a dual 6-form $C_6$ such that
\begin{equation}
\d C_6 = *_{10+1}\d C_3 = e^{-2\Phi}*_{9+1}H \; ,
\end{equation}
where the second Hodge-star operation is with respect to the string
metric as before. Of course, we may now identify $dB_6=dC_6$.
 Instead of setting $B_6=C_6$,
let us leave the additive gauge freedom in definition of $C_6$ for
a while.

According to Ref.\cite{Gutowski:1999tu}, an M5-brane with 4
translational symmetries along $x^\mu$ may be calibrated as follows.
The energy density functional per unit 3-volume is composed of two
pieces. The first is the warped volume density
\begin{equation}
\int_\Sigma\sqrt{-Det\left(e^{-2\Phi/3}\eta\right)}
\cdot \sqrt{Det\left(e^{-2\Phi/3} h\right)} \; ,
\end{equation}
with the induced metric $h$ from $g$ on the embedding
surface $\Sigma$, which is actually
\begin{equation}
\int_\Sigma e^{-2\Phi}\sqrt{Det\left( h\right)}
\end{equation}
The second piece is the Coulomb energy density, which goes as
\begin{equation}
-\int \Sigma^*\left( vol^\eta_{3+1}\lrcorner \;C_6\right) \; .
\end{equation}
The sum of these two terms are total energy density of the
configuration
\begin{equation}
{\cal E}'=\int_\Sigma e^{-2\Phi}\sqrt{Det\left( h\right)}
-\int \Sigma^*\left( vol^\eta_{3+1}\lrcorner \;C_6\right) \; .
\end{equation}
For calibration, there exists a closed 2-form $K$ such that
\begin{equation}
{\cal E}'\ge \int_\Sigma K \; ,
\end{equation}
is saturated precisely for the supersymmetrically wrapped branes.
The deformation of the surface $\Sigma \rightarrow \Sigma'$
would change the energy ${\cal E}'$ while
\begin{equation}
\int_{\Sigma'} K -\int_{\Sigma}K
=\int_{V}\d K=0\; ,
\end{equation}
with the interpolating volume $V$ whose two boundaries are
$\Sigma$ and $\Sigma'$. Thus, such a $K$, to be found via supersymmetry
conditions, will provide the absolute minimum energy to be
saturated.

The closed form $K$ is found as follows.
One first finds a covariantly constant spinor, $\epsilon$, which
is responsible for the supersymmetry of the background geometry.
The supercharge ${\cal Q}(\epsilon)$ associated with $\epsilon$
will have the property
\begin{equation}
2 {\cal Q}(\epsilon)^2={\cal E}'-\int_\Sigma K \; ,
\end{equation}
ensuring the generalized calibration. The 2-form $K$ is found
to be
\begin{equation}
K=-vol^\eta_{3+1}\lrcorner \;C_6+
\bar\epsilon\Gamma\epsilon\sqrt{-Det\left(e^{-2\Phi/3}\eta\right)} \; ,
\end{equation}
where $\Gamma$ is the pull-back of $\Gamma_{AB}\d X^A \d X^B$ to
$\Sigma$.
$\Gamma$ differs from its counterpart in the heterotic
string theory, call it $\gamma$, by an overall factor of
$e^{-2\Phi/3}$. Then the final form of $K$ is
\begin{equation}
K=-vol^\eta_{3+1}\lrcorner \;C_6+e^{-2\Phi}\,
\, \Sigma^*\left(\,\bar{\epsilon}\gamma_{mn}\epsilon\;\d x^m \d x^n\right) \; .
\end{equation}
The bulk 2-form in the latter term, we realize to be
precisely the fundamental 2-form $J$
of the internal dimension, so we have
\begin{equation}
K=-vol^\eta_{3+1}\lrcorner \;C_6+e^{-2\Phi}\Sigma^*(J) \; .
\end{equation}
$K$ should be closed, $\d K=0$, if we have supersymmetry in the
bulk, and indeed this follows from one of our supersymmetric
equation for any heterotic flux background, $e^{-2\Phi}*H=
\d \left(e^{-2\Phi}J\right)$. Finally the two $C_6$ pieces
in ${\cal E}'$ and $K$ cancel each other, and
\begin{equation}
{\cal E}= {\cal E}'-K=0 \; ,
\end{equation}
is the generalized calibration condition, regardless of the gauge
choice for $C_6$, as we promised above.

\section{A Smooth Compactification and Calibration}

A nonsingular flux compactification of the heterotic string theory
was recently found by Fu and Yau \cite{Fu:2006vj}, which was further
elaborated on by Becker et.al. \cite{Becker:2006et} The string
theory on this background is believed to be U-dual to an orientifold
limit of $K3\times K3$ compactification of F-theory with fluxes
turned on. One can describe the geometry as a $T^2$ fibred over a
conformally rescaled $K3$, where the metric and the fundamental
2-form are
\begin{equation}
g=e^{2\Phi}g_{K3}+|\theta|^2 \; ,
\end{equation}
and
\begin{equation}
J=e^{2\Phi}J_{K3}+\frac{i}{2}\,\theta\w \bar\theta \; ,
\end{equation}
with a holomorphic  1-form $\theta$. The dilaton is a
function on $K3$ only. Locally
$\theta$ should have the form
\begin{equation}\label{holo 1-form}
\theta=\d z+\alpha \; ,
\end{equation}
with ill-defined 1-form $\alpha$ on K3, where $z$ is the
holomorphic coordinate on the fibre $T^2$. Well-defined $T^2$
bundle over $K3$ requires that we have integral Chern classes,
which imposes that``real''and ``imaginary'' parts of the 2-form
\begin{equation}
\omega=\frac{\d\theta}{2\pi\sqrt{\alpha'}} \; ,
\end{equation}
belong to integral cohomology of $K3$. Since we must have
\begin{equation}
0=\d \left(e^{-2\Phi}J\w J\right)\quad\Rightarrow\quad
0=iJ_{K3}\w \left( \omega\w \bar{\theta}-
\theta\w\bar{\omega}\right) \; ,
\end{equation}
we must also require the primitivity of $\omega$ in $K3$.
\begin{equation}
\omega\w J_{K3}=0 \; .
\end{equation}
These exhaust constraints on the geometry except for determination
of the dilaton on $K3$ and supersymmetry condition on the gauge
bundle. Supersymmetry condition on the gauge bundle is the
familiar one. Namely, the field strength $F$ should be of type
$(1,1)$ and primitive with respect to $J$.

This ansatz solves all supersymmetry relationships except for
actual form of $\Phi$, where all functional information of the
solution is encoded. The authors of Ref.\cite{Fu:2006vj} choose to
extract the equation for $\Phi$ from the Bianchi identity,
\begin{align}
\d H = \frac{\alpha'}{4} \Big[ \tr ( R'  \w R' )
- \tr (F \w F) \Big]
\; ,
\end{align}
where they take the curvature 2-form $R'$ to be that of the
so-called Hermitian connection. This choice involves a torsion
which is not completely anti-symmetric and thus cannot be of the
form, $\sim a H$. Nevertheless, the size of this torsion is the
same order as $H$, and this choice represents a correction term of
order $\sim (\alpha')^2$ in this equation.\footnote{For smooth and
large flux compactification, it turns out that the size of $H^2$
has to be of order  $\alpha'$. See Ref.\cite{Kimura:2006af} for
more detailed explanation. This fact also holds in this solution,
naturally.} Also this has the advantage that the right hand side
is $(2,2)$ Hodge type, allowing the equation self-consistent
without a further $\alpha'$ truncation. We will come back to this
anomaly equation later on to discuss a crucial tadpole condition.

Before proceeding further, let us note that the sizes of
the base and of the fibre are free and also that the zero
mode of dilaton, $\Phi_0$, is free. A better way to write
the ansatz is,
\begin{equation}
g=e^{2(\Phi-\Phi_0)}R_B^2\hat g_{K3}+l_F^2|\theta|^2 \; ,
\end{equation}
and
\begin{equation}
J=e^{2(\Phi-\Phi_0)}R_B^2\hat J_{K3}+\frac{i l_F^2}{2}\,\theta\w \bar\theta \; ,
\end{equation}
where the volume of $K3$ in terms of $\hat g_{K3}$ is normalized
to unit. The linear size of the fibre is $2\pi l_F\sqrt{\alpha'}$
and the linear size of the base is $R_B$, both of which are free
parameters of the solution. The solution would be trustworthy when
$R_B^2/\alpha' \gg 1$ and $l_F\gg 1$. In the following
discussions, however, $l_F$ factor either cancels out or appears
as an overall coefficient while $e^{-\Phi_0}R_B$ can be absorbed
into the definition of the $K3$ metric as
$g_{K3}=e^{-2\Phi_0}R_B^2\hat g_{K3}$ and similarly of the
K\"ahler form. We will suppress these factors, with the
understanding that they can be restored easily and also that their
large sizes are important for our discussions.

This heterotic flux compactification is deemed to be U-dual to a
well-known F-theory compactification on $K3\times K3$, or more
precisely an orientifold limit thereof, provided that we add some
additional RR and NS-NS fluxes, $F_3$ and $H_3$, on the latter
\cite{Becker:2006et,Dasgupta:1999ss}. As
was shown by Sen \cite{Sen:1996vd}, F-theory on $K3\times K3$ \cite{Vafa:1996xn}
is a generalization of
the orientifold $T^2/Z_2$ times $K3$ of IIB theory, whereby we move
around the D7-branes located at the four tips of $T^2/Z_2$. In the
orientifold limit, each tip represents one $O7^-$ plane, 4 pairs of
$D7$ branes, and the associated $SO(8)$ gauge groups on D7's.

The duality chasing starts with T-dualization
of the $T^2/Z_2$. This will change $O7^- $ and $D7$'s
into $O9^-$ and $D9$'s, and the orbifold
is now resolved to regular $T^2$. However, the presence
of NS-NS flux $H_3$ means that certain off-diagonal value
of metric will be generated \cite{Buscher:1987qj}, so the geometry on this type I
side will be such that $T^2$ has nontrivial mixing with $K3$
in  the metric. Then, we can switch over to the
heterotic side by taking S-duality \cite{Witten:1995ex,Polchinski:1995df},
whereby mapping $F_3$ to $H$.

In this note, we are mostly interested in understanding possible
brane configurations on the two sides and comparing their
properties, in part because this will strengthen this duality
conjecture and also in part this could diversify possible models
based on such compactification. As we noted above, the IIB
background comes with D7 branes as part of the geometric data. On
top of this, we can think of two more classes of D-brane
configurations with supersymmetry. One class consists of  D3 branes
transverse to the compact directions. The other is D7 branes
intersecting with the indigenous D7 branes. The latter wraps $K3$
while the former wraps $T^2/Z_2$ and a 2-cycle on $K3$. The latter
would be supersymmetric only under a tight restriction on bulk
complex moduli, unlike the former. We will come back to moduli of
these D-branes in a later section.

Interestingly, both classes of these D-branes map to D5's in type
I, and then to fivebranes upon S-duality to the heterotic side. D3
can be seen to correspond to fivebrane wrapping the fibre $T^2$,
while D7's wraps 2-cycles in the base K3. For the rest of this
note, we will consider issues related to these calibrated
fivebranes on the heterotic side. A simplifying fact  is that the
presence of $e^{2\Phi}$ factor on the conformal rescaling of $K3$,
so that the Coulomb energy density has a simple form,
\begin{equation}
\int\Sigma^*\left(e^{-2\Phi}J\right)
=\int\Sigma^*\left(J_{K3}\right)+\frac{i}{2}
\int\Sigma^*\left(e^{-2\Phi}\theta\w\bar\theta\right) \; .
\end{equation}
Since $\d J_{K3} = 0$,
the first piece in the right-hand-side is topological
\begin{equation}
\delta\int \Sigma^*(J_{K3})=0 \; ,
\end{equation}
so the problem of minimizing the energy functional becomes
that of saturating the bound
\begin{equation}\label{bound}
{\cal E}'=\int_{\Sigma}e^{-2\Phi}vol_\Sigma-\frac{i}{2}\int
\Sigma^*\left(e^{-2\Phi}\theta\w\bar\theta\right) \ge
\int \Sigma^*(J_{K3}) \; .
\end{equation}
In terms of the generalized calibration presented in the
previous section, this split corresponds to the choice of
gauge for $C_6$,
\begin{equation}
C_6=\frac{i}{2}\,e^{-2\Phi}\theta\w\bar\theta\w vol^\eta_{3+1} \; ,
\end{equation}
forcing
\begin{equation}
K=J_{K3}\;,
\end{equation}
which is indeed a closed 2-form. In the large volume limit,
${\cal E}'$ is a well-normalized measure of the energy density
associated with the fivebrane.

The configuration $\Sigma$ saturating the lower bound
$\mathcal{E}'(\Sigma)=\int\Sigma^*(K)$ or equivalently
${\cal E}=0$ satisfies the local condition
\begin{equation}
  vol_\Sigma=\Sigma^\ast(J)\ .
\end{equation}
Since $J$ is a Hermitian (1,1) form on $\mathcal{M}_6$, this
condition is satisfied if the embedding of $\Sigma$ into
$\mathcal{M}_6$ is holomorphic. In the background geometry
described in section 3, two independent tangent vectors of
$\Sigma$ pushed forward into the bulk may be written as
\begin{equation}
  \xi+a\partial_z\ ,\ \ \bar{\xi}+\bar{a}\partial_{\bar{z}} \; ,
\end{equation}
with suitable coefficient $a$. Here $\xi$ is a tangent vector in
$K3$, which is $(1,0)$ with the complex structure $J_{K3}$. Since
$\mathcal{M}_6$ is a bundle over the base $K3$, let
us call the bundle projection map, $\pi$,
\begin{equation}
  \pi:\ \mathcal{M}_6\rightarrow K3\ .
\end{equation}
One can  see that the image $\pi(\Sigma)$($\subset K3$) of
the holomorphic embedding $\Sigma$ is also holomorphic, with
tangent vectors $\xi$ and $\bar{\xi}$. The holomorphic surface
$\pi(\Sigma)$ determines the value of the integral at the right
hand side of (\ref{bound}). Thus,
we can think of two distinct classes of solutions, depending
on whether the pull-back of $J_{K3}$ integrates to zero or not.

\section{Fivebranes on $T^2$, Cyclic Homotopy,
a Tadpole, and U-Dual D3 Branes}

We first consider the case
\begin{equation}\label{vanishing}
  \int \Sigma^\ast(J_{K3})=\int(\pi(\Sigma))^*(J_{K3})=0\ .
\end{equation}
A holomorphic embedding $\pi(\Sigma)$ in $K3$ with vanishing
integral (\ref{vanishing}) is necessarily point-like. Therefore,
$\Sigma$ can wrap only the $T^2$ fibre  and is localized at the
point $\pi(\Sigma)$ in the base. We may take the complex
coordinate $\zeta$ on the internal part of the worldvolume
embedded as
\begin{equation}
  z=\zeta\
\end{equation}
where $z$ is the complex coordinate of the fiber introduced in
section 2, with identifications $\zeta\sim\zeta +2\pi
m\sqrt{\alpha'}$ and $\zeta\sim\zeta+2\pi n i\sqrt{\alpha'}$ ($m$,
$n$ are integers) since $\Sigma$ is topologically a torus now.
Note that the saturation (\ref{vanishing}) of
$\mathcal{E}(\Sigma)\geq 0$,
\begin{equation}
  \Sigma^\ast(e^{-2\Phi}J)=
  \frac{i}{2}\Sigma^\ast(e^{-2\Phi}dz\wedge d\bar{z})
  =e^{-2\Phi}vol_\Sigma\ ,
\end{equation}
occurs automatically for a completely vertical configuration. With
this, we have ${\cal E}'(\Sigma)=0$.

An interesting fact about the fibre $T^2$ is that it does not
correspond to an element of the homology group with real
coefficients.
The effect of twisting due to $\omega_i$ is that the two
circles of $T^2$ can become a contractible loop. As a toy
example, take an $S^1$ fibred over $S^2$ via Hopf fibration.
The metric goes as
\begin{equation}
\d s^2_{S^2}+(\d\psi+k\cos\theta\d\phi)^2 \; ,
\end{equation}
with the Hopf number $k$. The resulting topology of the bundle
is $S^3/Z_k$ and its first homotopy group is
\begin{equation}
\pi_1=Z_k  \; .
\end{equation}
A loop that winds around the fibre $S^1$ $k$-times becomes
contractible. The twisting in ${\cal M}_6$ is essentially the same
type of fibration, except that we now have a pair of $S^1$'s and
that the two-dimensional bases are replaced by 2-cycles in $K3$.
Because of this, a fivebrane wrapping $T^2$ $(m,n)$-times will be
homotopically trivial when $m$ or $n$ equals some integer $k$ whose
precise value is determined by the bundle ${\cal M}_6$. Because a
fivebrane can unwrap when either of the circle becomes contractible,
there are in general at least two such integers $k_1$ and $k_2$,
associated with the two circles of $T^2$ fibre, so that we have in
general processes that shifts
\begin{equation}
\Delta N_{fivebranes\; on \;T^2}=a_1k_1+a_2k_2
\end{equation}
for integers $a_1$ and $a_2$.

A simple generalization of this toy model is to consider $S^1$ fibred
over $(S^2)^K$, with  the metric being
\begin{equation}
\sum_{p=1}^K\d s^2_{S^2_p}+
\left(\d\psi+\sum_pm_p\cos\theta_p\d\phi_p\right)^2 \; ,
\end{equation}
in which case the homotopy group would be $Z_k$ with $k$ being
the greatest common divisor of $\{m_p\}$. Although in our case the
1-forms that enter the last terms are associated with 2-cycles in
a single $K3$,  this still suggests that
$k_{1,2}$ above should be determined similarly by what integral
linear combination of the generators $H^2(K3,Z)$ is used for
$\omega_{1,2}$.

However, this does not mean that the dominant
unwrapping processes are the ones that shift the winding number
by $k_1$ and $k_2$. Depending on precise geometry, some linear
combination like $2k_1+5k_2$ could prove to be the easiest path.
Homotopy does not know anything about the dynamics. In terms of
the toy model geometry, the ``easiest" path would correspond to
shift of the winding number by $a_1k_1+a_2k_2=m_p$ if $p$-th $S^2$
happens to be much smaller than all the others.

In any case, with such a cyclic nature of the winding number,
our finding that fivebrane wrapping $T^2$ is stable and supersymmetric, may sound strange.
Wrapping it $k$-times will result in homotopically trivial
configuration which can be unraveled and made to contract to
nothing. In fact, depending on details of the metric, it may
even be possible to deform $\Sigma$ away from this vertical
configuration and
reduce its area. The point is that such a deformation is
always accompanied by a cost in the Coulomb energy,
\begin{equation}
\Delta E_{electric}=-\int_{\Sigma'}
\,\frac{i}{2}\,e^{-2\Phi}\theta\w\bar\theta
+\int_{\Sigma} \,\frac{i}{2}\,e^{-2\Phi}\theta\w\bar\theta \; ,
\end{equation}
and that this cost always
override, if any, the energy gain from the reduction of
the area when the fivebrane is deformed away from $T^2$.
It is the magnetic Coulomb energy that protects such
nontopological configurations.\footnote{
While the energy density ${\cal E}'$ is such that the
fivebrane winding the $T^2$ fibre has the same energy as the
trivial configuration, one should not take this mean that
the $T^2$  wrapped fivebrane is tensionless. The coupling
to the Coulomb field does not enter the dynamics of the
general worldvolume excitations.}

We just argued that the winding number of fivebrane over the fibre
$T^2$ is additive only modulo some integers $k_i$'s.
Consistent with this fact is that there is no element of
$H^2({\cal M}_6,R)$ which is Poincare-dual to the $T^2$ fibre
\cite{Goldstein:2002pg,Becker:2006et}.
This non-extensive or cyclic nature of $T^2$ winding number
raises a number of interesting questions.

First, recall that fivebranes contribute to the tadpole
condition for $H$ since it is a magnetic source. The $T^2$ winding
number will act as a magnetic source to $H$ field along $K3$.
How is it possible that such a tadpole source can be unwrapped
and disappear?
The resolution to this quandary comes from the fact that $K3$
is not a cycle in the manifold ${\cal M}_6$. The fivebrane
in question wraps the fibre $T^2$, and therefore can
contribute to a tadpole condition along the base $K3$.
Normally this tadpole condition would arise from integrating
$\d H$ equation over a homology cycle representing $K3$.
However, there is no such cycle as long as the $T^2$ bundle is
nontrivial. Of course, in the present geometry, we can unwrap
the fivebrane on fibre precisely when the $T^2$ bundle is
nontrivial.

Instead, the relevant tadpole condition arises from
integration of $J\w \d H=\cdots$ over the entire manifold,
\begin{equation}
\int_{{\cal M}_6}J\wedge \d H=\frac{\alpha'}{4}\left[\int_{{\cal M}_6}
J\wedge\left(tr R'\wedge R' - tr F\wedge F-16\pi^2\delta_{fivebrane}\right)\right] \; .
\end{equation}
Note that the left hand side does not vanish. Instead
we have
\begin{equation}
\int_{{\cal M}_6} J\wedge \d H =4\pi^2\alpha'\int_{K3}|\omega|^2
\; ,
\end{equation}
where the contraction of $\omega$ is taken with respect to the
Calabi-Yau metric $g_{K3}$ without the dilaton factor.
The right hand side can also be computed, and
we have the following  tadpole condition
\begin{equation}
N_{fivebrane\; on\; T^2}+
\int_{K3}|\omega|^2=\frac{1}{16\pi^2}\int_{K3} tr R_{K3}\wedge R_{K3}-tr F\wedge F \; .
\end{equation}
The right-hand-side is
\begin{equation}
-\frac12 p_1(K3)+\frac12 p_1({\cal F})
\end{equation}
in terms of the Pontryagin class $p_1$. We have $-p_1(K3)/2=24$
while $-p_1({\cal F})/2$ is a nonnegative integer. Note that this
equation does reflect the expectation that the fivebrane number is
interchangeable with the topological number on the right hand side.

Now consider the process of unwrapping a fivebrane on $T^2$ with all
backreaction taken into account. This process will necessarily
involve a fivebrane source which is spread over the base $K3$ as
well as over $T^2$. The backreaction of the metric and the torsion
is then such that this fibre-bundle form of the geometry is
completely destroyed throughout the middle step. Since we are not
resolving the fivebrane into the gauge bundle, neither of the
Pontryagin numbers will change upon completion of this unwrapping
process, leading to our quandary.

However, the second term on the left, which has something to do with
the winding of $T^2$ fibre can easily change. The initial and the
final configurations are $T^2$ bundles over $K3$, but the
interpolating geometries cannot be one. This deformation of the
bundle structure should absorb the difference in $N_{fivebrane\;on\;
T^2}$ and encode it in the integer shift of the integral of
$|\omega|^2$. This could occur via a shift of the integral
cohomology associated with $\omega_{1,2}$, or due to change of
metric on the base $K3$. In the former case, another $T^2$, which is
different from the initial $T^2$ fibre, emerges as the fibre at the
end of the process. In the latter case, the cohomology class of
$\omega_{1,2}$ remains unchanged but their split into self-dual and
anti-self-dual part can change. Because the intersection pairing
(relevant to cohomology class) and the norm differ by a sign in
anti-self-dual part of $H^2(K3)$, this can also shift the integrated
value of $|\omega|^2$.

This does not mean that a homotopically trivial fivebrane can decay
to nothing. The winding configuration represents a supersymmetric
vacuum in 3+1 dimensional effective theory, and thus cannot decay to
another supersymmetric vacua. It only means that we have two (or
more) degenerate configurations with different $T^2$ winding numbers
of the fivebrane. In particular, there should be supersymmetric
domain wall configurations separating these degenerate vacua from
one another. It is not difficult to see that the domain walls
themselves should be represented by fivebranes wrapping a 2-cycle in
$K3$ and one of the two circles in $T^2$. Following how
configuration changes as we move from the side with winding to the
other side, one should see the fivebranes gradually unwinding as a
function of the transverse coordinate.

So far, we addressed various issues entirely within the heterotic
theory. Here we would like to close the section by studying how this
unusual nature of $T^2$ wrapped fivebranes manifest itself in the
U-dual picture. As we noted earlier, these fivebranes map to D3
branes in IIB orientifold. Seemingly, D3 branes carry an integer
conserved charge, and we have a potential conflict.

This can be made  more dramatic by replacing D3 by anti-D3. In the
heterotic side, one wraps the fivebrane over $T^2$ with opposite
orientation. Since this breaks supersymmetry, the false vacuum will
try to decay into a supersymmetric one taking away $k$ unit of
anti-fivebranes.  Anti-fivebranes on $T^2$ have an energy function
which is just twice the tension and would favor being unwrapped out
of $T^2$. Without topology protecting them, a $k$ number of
anti-fivebranes on $T^2$ can then disappear either classically or by
tunneling. Can anti-D3's also disappear in some quantized unit on
IIB side? Although we just phrased the question in terms of
anti-D3's to make the possible conflict more obvious, the same sort
of question exists for D3's as well. In term of the latter, the
question is whether there are supersymmetric domain wall
configurations with different number of D3 branes on the two sides.

On IIB side, D3 tadpole condition involves a flux contribution
so that we have
\begin{equation}
0=\int F_3\wedge H_3 +N_{D3}
\end{equation}
where appropriate normalization constants are understood. On the
other hand, $F_3$ and $H_3$ fluxes can jump in quantized unit across
a domain wall formed by a D5-brane or by a NS5-brane wrapping a
3-cycle \cite{Gukov:1999ya}. The tadpole must be preserved no matter
what, and $N_{D3}$ will thus jump across such a domain wall.

When one side of this domain wall contains anti-D3 branes, the
domain wall can form a bubble, inducing decay of a false vacuum with
anti-D3 into another false vacuum with lesser number of anti-D3 or
into a true vacuum with no anti-D3 branes. The precise unit in which
$N_{D3}$ jumps depends on what are initial fluxes in $F_3$ and
$H_3$, and which 3-cycle is used for wrapping NS5 or D5 branes to
form the domain wall. If $H_3$ flux is shifted by unit via a domain
wall from NS5 wrapping a cycle $A$, $N_{D3}$ will shift by an
integer
\begin{equation}
\oint_B F_3
\end{equation}
where $B$ is the dual 3-cycle of $A$, and vice versa. Set of
these integers, that is, possible shifts of $N_{D3}$ via
such domain walls, should match precisely the allowed linear
combinations of the two integers $k_{1,2}$'s we
saw in the heterotic side, if the U-duality holds in the
presence of the flux. As in the heterotic case, if one side
contains anti-D3 branes, breaking supersymmetry, the wrapped
D5 or NS5 branes will instead appear as an expanding bubble
in 3+1 dimensions, removing anti-D3 branes in these quantized
units.

Finally, it is not entirely
obvious why the unwrapping of anti-fivebranes has to be via
tunneling on the heterotic side. Of course the U-dual map involves a strong-weak
coupling duality, so we cannot compare directly. Still, ignoring
quantum issues, it looks likely that, even in the heterotic side,
the decay of anti-fivebrane on $T^2$ would occur via tunneling.
Taking T-dual of a large IIB orientifold along the $T^2/Z_2$, the
base manifold remains large while the fibre may not. An important
fact is that the ratio of the size of the base and that of the
fibre could be a large number, and this is unaffected by the final
S-duality into the heterotic side. Furthermore, anti-fivebrane on
$T^2$ will settle down to a point where $e^{-2\Phi}$ is minimized
along $K3$, where the conformally rescaled size of $K3$ is
maximal. These two act together to force the unwrapping process to
be a tunneling process, if the fibre size is much smaller.
Unwinding necessarily involves fivebrane wandering into $K3$,
which could come at the cost of much larger area. Recall that for
anti-fivebrane the net energy is bounded below by the tension
energy due to the area, since the wrong sign of the Coulombic
energy term adds rather than subtract. Also, being at local
minimum of $e^{-2\Phi}$ implies that unwrapping will cost even
more because the configuration must move away from the local
minimum of this factor also.

There are still more details of this matching between $T^2$ wrapping
fivebranes and D3 branes that must be
checked. In particular, it remains  a challenging problem to match
of all possible tunneling
processes on F/IIB side which changes the number of D3 branes
against those in the heterotic side which changes the $T^2$ winding number.
Equivalently, one would like to match all possible domains walls
of the two sides.
However, a precise matching of this kind
would go a long way in establishing the duality map we have
been using and should be a worthwhile exercise.

\section{Fivebranes on 2-Cycles in $K3$}

Next, we consider the case
\begin{equation}\label{non-vanishing}
  \int\Sigma^\ast(J_{K3})\neq 0\ .
\end{equation}
In this case, $\pi(\Sigma)$ spans a two dimensional surface,
and represents a nontrivial
element in the second homology $H_2(K3)$ group. To
saturate the calibration bound condition,
$\pi(\Sigma)$ must be a holomorphic
embedding in $K3$, which means that the pull-back of the
holomorphic (2,0)-form of $K3$ vanishes
\begin{equation}
0=\pi(\Sigma)^*(\Omega^{(2,0)}_{K3}) \;.
\end{equation}
Once we find such a holomorphic
embedding $\pi(\Sigma)$, we should search for a uplift
it to the total manifold ${\cal M}_6$.

This uplifting is not possible unless
there is a global section. That is, this last step is
possible only if the restriction
of the bundle over $\pi(\Sigma)$ is a trivial bundle.
This in turns requires that the Chern class of the bundle
integrates to zero over this surface, so that we must have
\begin{equation}\label{orthogonal-1}
  \int_\Sigma d\alpha=0\ \ \rightarrow\ \ \
 \int_\Sigma \omega_1=\int_\Sigma \omega_2=0\ ,
 \end{equation}
as well. Once this holds, it comes down to what kind
of uplifting is available, which will depend on topology
of $\pi(\Sigma)$.

Because the restricted bundle
$\pi^{-1}\left(\pi(\Sigma)\right)\subset {\cal M}_6$ is trivial,
we can also talk about $T^2$-winding number of this uplifting.
That is, $\Sigma$ may wind around the fibre once or more,
depending on the topology of $\pi(\Sigma)$, and this would
provide additional quantum number associated with $\Sigma$.
Note that this $T^2$ winding number does not contribute to
${\cal E}'$ since the contribution to the area is canceled
by the Coulomb energy point-wise. Thus, the uplifting of
$\pi(\Sigma)$ with additional $T^2$ winding number represents
a threshold bound state of $\Sigma$ without the $T^2$ winding number
and a number of the $T^2$ wrapping fivebrane.

While the above more or less characterize solutions to the
supersymmetry conditions, we can do things more explicitly
thanks to the well-known description of the
homology of $K3$. We can translate all of
above as a set of restrictions on various intersection
numbers.  Let us call
the generators of this homology $\Sigma_I$
($I=1,\cdots,22$), whose intersection numbers are given by the
matrix
\begin{equation}
  C_{IJ}=\left(\begin{array}{ccccc}
  -E_8&&&&\\&-E_8&&&\\&&U&&\\&&&U&\\&&&&U\end{array}\right)\ \ ,
  \ \ \ U=\left(\begin{array}{cc}0&1\\1&0\end{array}\right) \; ,
\end{equation}
where $E_8$ denotes the $8\times 8$ Cartan matrix of the $E_8$ Lie
algebra. We expand
\begin{equation}\label{cycle expand}
  [\pi(\Sigma)]=\sum_{I=1}^{22}n^I[\Sigma_I] \; ,
\end{equation}
where $n^I$ are integers. The 2-forms $\eta_I$ Poincare-dual to
$\Sigma_I$ are defined by the relation
\begin{equation}
  \int_{K3}\eta_I\wedge v=\int_{\Sigma_I}v \;,
\end{equation}
for any closed 2-form $v$.

$J_{K3}$ is covariantly constant in terms of $K3$ metric,
and can be expanded as
\begin{equation}
  J_{K3}=\sum_{I=1}^{22}h^I \eta_I \; ,
\end{equation}
with real $h^I$. $K3$ is Calabi-Yau, or equivalently,
hyperK\"ahler, and has a holomorphic 2-form $\Omega^{(2,0)}_{K3}
$, which can also be represented as a linear combination of
$\eta$'s,
\begin{equation}
  \Omega^{(2,0)}_{K3}=\sum_{I=1}^{22}t^I\eta_I\ .
\end{equation}
The pull-back of $\Omega^{(2,0)}$ must vanish on $\pi(\Sigma)$,
so we have
the first necessary condition for calibrated $\pi(\Sigma)$
\begin{equation}
\int_{\pi(\Sigma)}\Omega^{(2,0)}_{K3}=\sum_{IJ}C_{IJ} n^It^J =0 \; .
\end{equation}
Furthermore, the surface should have the right orientation
to have positive integral of $J_{K3}$ which demands that
\begin{equation}
  \int_\Sigma\Sigma^\ast(J_{K3})=n_I h^J\int_{\Sigma_I} \eta_J=
  C_{IJ}n^I h^J > 0\ .
\end{equation}
Obviously at least one of $n^I$ should be nonvanishing.

Given such a holomorphic embedding $\pi(\Sigma)$, there is
well-known counting of its moduli in $K3$. Thanks to $\Omega^{(2,0)}$,
counting of deformation become counting of $H^1(\Sigma)$ \cite{Bershadsky:1995qy},
which is in turn related to the Euler number as
\begin{equation}
\hbox{dim}\, H^1(\Sigma)=2-\chi =2g \; .
\end{equation}
There are in general $2g$ real deformation moduli
for genus $g$ surface $\pi(\Sigma)$. In terms of the above
decompositions into integral cohomology, there is an alternate
formula for this number, which goes as
\begin{equation}\label{euler}
2g\stackrel{\star}{=}2-\int_\Sigma c_1(T_{\pi(\Sigma)})
  \stackrel{\checkmark}{=}2+ \int_\Sigma c_1(N_{\pi(\Sigma)})=
2+\int_{K3}\sigma\wedge\sigma
  =2+C_{IJ}n^I n^J
\end{equation}
where at $\checkmark$ we used the fact
$0\!=\!c_1(K3)\!=\!c_1(T)\!+\!c_1(N)$, and at $\star$ we assumed
that the surface is \textit{connected} and \textit{smooth}. The
$\sigma$ in the integrand denotes the Poincare dual of the 2-cycle
$\pi(\Sigma)$, $\sigma=\sum_I n^I\eta_I$. Existence of such a
holomorphic embedding implies that the deformation parameter is
nonnegative, so we arrive at
\begin{equation}
C_{IJ}n^I n^J \ge -2
\end{equation}
which is the 3rd necessary condition.

So far we considered the implication of having nontrivial
holomorphic image $\pi(\Sigma)$ in the $K3$ base. As we saw above,
uplifting this to a holomorphic surface in ${\cal M}_6$ requires
$\pi(\Sigma)$ be orthogonal to $\omega_{1,2}$ under the
intersection pairing. Writing
\begin{equation}
  \omega_i=\sum_{I=1}^{22}l_i^I\eta_I\ \ \ \
  (i=1,2\ ,\ \ l_i^I\ {\rm are\ integers})\ .
\end{equation}
we have the 4th necessary condition
(\ref{orthogonal-1}) as
\begin{equation}\label{orthogonal-2}
  C_{IJ} n^Il_i^J=0\ ,
\end{equation}
When this holds a global section exists, and we could uplift
$\pi(\Sigma)$ into ${\cal M}_6$.

Summarizing, we have several necessary conditions among the
homology cycle $\pi(\Sigma)=n_I[\Sigma_I]$, $J_{K3}=h^I\eta_I$,
$\Omega^{(2,0)}_{K3}=t^I\eta_I$,
and $\omega_i= l_i^I\eta_I$
\begin{itemize}
\item $C_{IJ} n^I t^J =0$
\item $C_{IJ} n^I h^J > 0$
\item $C_{IJ} n^I n^J \ge -2$
\item $C_{IJ} n^I l_i^J =0$
\end{itemize}
in addition to the bulk supersymmetry conditions
\begin{itemize}
\item $C_{IJ} h^I t^J = 0$
\item $C_{IJ} t^I t^J = 0$
\item $C_{IJ} h^Il_i^J =0$
\end{itemize}
all of which must hold in order for the supersymmetric $\Sigma$ to
exists.

There could be at most 20 independent integer vectors $n^I$'s
which solve the these equations. The precise number of which
depends on whether $\omega_i$'s are entirely along $(2,0)$ or has
component in $H^{(1,1)}$. Even for those solution that solves all
of above requirement, many of them cannot be realized as a single
holomorphic surface of the right orientation. As a crude example,
working with a 2 dimensional subspace, we take $\omega_1=(1,\pm
1)$, $\omega_2=(0,0)$ with
\begin{equation}
  C_{IJ}\sim\left(\begin{array}{cc}2&0\\0& 2\end{array}\right)\ .
\end{equation}
The solution $u^I$ satisfying $C_{IJ}\omega_i^I n^J=0$ is
$n^I=(1,\mp 1)$ in each case. The upper sign should be forbidden
since it has wrong orientation.

Another set of interesting example where solution to the
above algebraic equations does not guarantee actual holomorphic
surface can be found when we confine ourselves to generators of
one of $E_8$ factor in the $H_2(K3)$. Here we expect only those
surfaces which actually exists as supersymmetric state are those
corresponding to a root of $E_8$. Because
$-C_{IJ}n^In^J$ measures the length squared of the corresponding
root, all 2-cycles of this kind are spheres with $g=0$. Other
combinations such as $\alpha+2\beta$, where $\alpha$ and $\beta$ are
pair of distinct roots, may be arranged to solve all of above constraints
but cannot corresponds to an irreducible, smooth, connected,
and holomorphic surface.

Reversely, the existence of such holomorphic $\pi(\Sigma)$
imposes constraints on the base $K3$ manifold. Each
wrapped 2-cycle of this kind must belong to the set
\begin{equation}\label{picard}
  {\rm Pic}(K3)\equiv H^2(K3,\mathbb{Z})\cap H^{1,1}(K3)
\end{equation}
which is called the Picard lattice. The Picard lattice is null
for generic $K3$, while the rank of this lattice can be as larger
as $20$ by adjusting $\Omega^{(2,0)}_{K3}$ since $h^{1,1}=20$.
When the rank is maximal the corresponding $K3$ is called
{\it attractive} \cite{Aspinwall:2005ad}.
This picks out $\Omega^{(2,0)}_{K3}$ among
discrete possibilities, so the remaining geometric moduli
are all in the choice of $J_{K3}$. Since $J_{K3}$ should be
orthogonal to $\Omega^{(2,0)}_{K3}$, an attractive $K3$ would
have 20 moduli intact out of the original 58 $K3$ moduli.

However, this statement should be taken
with a grain salt. Here we are pretending that we could
ignore back reaction of the geometry to such wrapped
fivebranes, and in particular are yet to take into account
the modification of tadpole conditions due to these new
sources. We believe consistent treatment of these effect
will take us away from the $T^2$ fibre bundle over $K3$,
and will require
much more elaborate geometry. The above statement of further
fixing of bulk moduli should be taken verbatim in
noncompact local models only, and for compact cases, backreaction
of the geometry should be taken into account. On the other
hand, having these extra fivebranes essentially generates
further fluxes, either in the form of gauge bundle or $H$ itself.
The idea of backreaction to additional fivebranes fixing
more moduli must be robust. We wish to come back to this
issue in a separate work.

\section{More on U-Dual and D-Branes}

U-dualizing to IIB orientifold $T^2/Z_2\times K3$ involves an
S-duality to type I and then T-duality on $T^2$. Following the
usual rules, it is quite clear that these fivebranes on $T^2$
are U-dual to D3-branes in IIB side. Similarly fivebranes wrapping
2-cycles in $K3$ are D7-branes, in IIB side, which wraps $T^2/Z_2$
as well as the same 2-cycles in $K3$. Here let us address further
issues related to this U-duality map between branes, such as
the complete 3+1 dimensional massless spectra of the two sides.

Before proceeding, we must first clarify which heterotic theory we
are considering. Since we have a $T^2$ in the geometry,
U-dualization can bring us to either of the pair of the heterotic
theory with $E_8\times E_8$ or $SO(32)$, depending on whether we
perform one more T-duality on one of the circle in the fibre or
not. Also the supersymmetry requirement solved by the current
solution is common to both theories, so it remains ambiguous which
theory we are discussing. For us, the main difference is in
fivebranes. Fivebranes in $E_8\times E_8$ theory are direct
descendants of M5 branes, and are equipped with a tensor theory on
worldvolume. Fivebranes wrapped on a circle maps under the T-dual
map to fivebranes wrapped on a dual circle in $SO(32)$ heterotic
theory, so the worldvolume theory on the latter must be a vector
theory \cite{Witten:1995gx}.

On the other hand, T-dual of a transverse fivebrane
is instead a KK monopole solution. Thus, when we identify the fivebranes
on a 2-cycle in $K3$ as U-dual of D7 branes wrapping $T^2/Z_2$ and a
2-cycle in $K3$, we are implicitly considering them in the heterotic theory
which is S-dual to type I, without further T-dualization on the fibre,
and this is $SO(32)$ heterotic theory. Thus, when we compare fivebranes
in the heterotic side to D7 on IIB side, we are considering $SO(32)$
theory and fivebranes whose worldvolume theory is a vector theory.

With this said, the comparison of the spectra is well established
in the absence of flux. Let us consider the low energy spectra of
fivebranes wrapped on 2-cycle $\Sigma$ and those of D7 branes
wrapped on $T^2/Z_2$ and $\Sigma$. Both carry a vector theory, and
the only difference come from how $T^2$ worth of position moduli
of the fivebrane arise from Wilson lines on D7 side. In addition,
there are $g$ complex moduli with $g$ being the genus of $\Sigma$
and a vector on both sides. Similar consideration shows identical
spectra for D3 and its U-dual. Generic D3 has 6 translational
moduli and one 3+1 dimensional vector. These are also obvious from
fivebranes on $T^2$. For each  fivebrane on $T^2$, there are four
translational degrees of freedom for a position on $K3$. Four more
bosonic massless degrees of freedom arise from the worldvolume
vector multiplet (or the self-dual tensor multiplet) on the
worldvolume, from two Wilson lines generating two scalars in
addition to the vector field itself.\footnote{If we were dealing
with fivebranes from $E_8\times E_8$ side, the worldvolume has a
tensor multiplet instead. But the same 3+1 dimensional spectra
arises since a 5+1 dimensional supersymmetric tensor theory
compactified on \ a circle gives the same field content as a 5+1
dimensional supersymmetric vector theory.} On IIB side, also
without the fluxes, the same counting appears from the adjoint
sector of D3 multiplet. The latter corresponds to a Coulomb phase
massless degrees of freedom of a single $N=2$ $D=4$ $SU(2)$.

What may be less obvious is what happens to this correspondence
when fluxes are turned on. For instance, it is well-known that
$D3$'s are attracted to D7's in the presences of NS-NS $B$-field
along directions transverse to D3 and longitudinal to D7. This can
be understood from the worldvolume theory of D3 as a
Fayet-Illiopoulos D-term constants on D3, which lifts its Coulomb
phase to have a finite and positive energy. This effect comes from
anti-self-dual part of  $B^{NS-NS}$ along $K3$. D3 would then
become a pair of non-commutative instantons in D7 gauge theory.
U-dual map of this configuration is likely to be a gauge bundle
over $K3$ on the heterotic side, so this would imply that isolated
fivebranes wrapping $T^2$ is not really there. Conversely, this
lifting of those moduli of D3 along $T^2/Z_2$ would map on the
heterotic side to massive Wilson lines on $T^2$, which seems very
unlikely.

In this case, the problem is solved
because of the detailed form of the NS-NS $B$-field. The twisting
of the bundle, encoded in the holomorphic 1-form $\alpha$ on $K3$,
arises from T-dualization of NS-NS $B$ form on IIB side. The
gauge choice which is convenient for the T-dualization is
\begin{equation}
B^{NS-NS}\sim df\wedge\alpha
\end{equation}
with $f$ denoting some function on $T^2/Z_2$. Note that the pull-back
of this to D7 wrapping $K3$ is identically zero, and that the effective
FI constants on D3 near the D7 vanishes. Even if we chose another gauge
such as
\begin{equation}
B^{NS-NS}\sim f\wedge\d\alpha
\end{equation}
$B^{NS-NS}$ is odd under the orientifold projection and $f$ has to
vanish at the four fixed points of $T^2/Z_2$, and in this orientifold
limit the D7 branes are precisely located at these fixed points.
Again the FI constants are not turned on D3 near D7, and the Coulomb
branch moduli of D3's are intact.

Precise matching of moduli of branes wrapping 2-cycles in $K3$ is a lot
more involved problem, in part because on the IIB side D7 branes in
question could have a worldvolume gauge bundle \cite{Gomis:2005wc}.
For this, we need a similar characterization of supersymmetric conditions
on the heterotic side as well. General supersymmetry condition on D-branes
in flux compactification of type IIB theory is by now understood fairly well
\cite{Martucci:2005ht,Martucci:2006ij}.
It would be most interesting to translate this to the heterotic side
and formulate the most general supersymmetric condition including the
worldvolume field strengths in the presence of bulk flux.

\section{Summary}

We studied calibrated fivebranes in flux compactification of the
heterotic string theory. Internal $H$-flux induces magnetic Coulomb
potential for the internal part of the worldvolume, and a consistent
supersymmetry condition in the absence of worldvolume field strength
is found. The resulting calibration condition can have nontrivial
solutions even without a topological winding number in a manner
consistent with tadpole conditions. We apply this setup to a recent
class of solution based on $T^2$ fibre-bundle over a conformally
rescaled $K3$, and found two distinct set of supersymmetric
fivebranes. The fivebranes wrapping $T^2$ are supported by flux and
also by a cyclic homotopy only, while the other is conventionally
calibrated with respect to a K\"ahler form of $K3$.

We also considered U-dual of this heterotic theory, realized as
IIB orientifold $T^2/Z_2\times K3$. Two types of calibrated fivebranes
are found to be dual to D3 branes and certain D7 branes, respectively,
and we matched some simple properties of the two sides.
In particular, we noted how the non-extensive nature of
$T^2$ winding number is mimicked by D3 branes in flux
compactification of IIB theory, and performed a qualitative
analysis.

Obviously this study still leaves much unaddressed. One
interesting extension would involve precisely matching domain
walls in IIB theory and those in the heterotic theory. In the
latter, the domain wall interpolating two regions with different
$T^2$ winding number of fivebranes, should be realized as a smooth
configuration of fivebranes themselves. This is different from IIB
side, where one find D3 branes ending on D5 or NS5.

Another interesting followup study would be how the low energy
effective theory would look like in the presence of fivebranes wrapping the
base $K3$. For these, it is important to understand the worldvolume
moduli fields better. However, this could turn out to be a difficult task
since for consistency we should be careful to take into account
the full effect of backreaction, as far as compact manifold goes.

\vskip 2cm
\centerline{\bf\large Acknowledgement}
\vskip 1cm

\noindent
We are indebted to  Li-Sheng Tseng and Alessandro Tomasiello for
useful comments, and also grateful to Tetsuji Kimura and Qing-Guo Huang
for discussion. PY would like to thank USTC Shanghai Institute
for Advanced Study, and the Center for Mathematical Physics for
hospitality, and in particular to ICTP and the organizers of ICTP
Workshop "String Vacua and Landscape" where much of the current
work was done. PY was supported in part by the Science Research Center Program of
the Korea Science and Engineering Foundation through
the Center for Quantum Spacetime(CQUeST) of
Sogang University with grant number R11-2005-021.
SK was supported in part by KOSEF Grant R010-2003-000-10391-0.

\vskip 2cm


}
\end{document}